\documentclass{article}
\usepackage{amssymb}
\usepackage{amsfonts}
\begin{document}

\begin{center}
{\LARGE {\bf Nonrenormalization theorems for  }}

\vspace{2mm}

{\LARGE {\bf $N=2$ Super Yang-Mills }}
 
\vspace{5mm} 

{\large V.E.R. Lemes$^{{\rm (a) }}$, N. Maggiore$^{{\rm (b) }}$, 
M.S. Sarandy$^{{\rm (c)}}$, S.P. Sorella$^{{\rm (a) }}$, \\ A. Tanzini$^{{\rm (d) }}$ and 
O.S. Ventura$^{{\rm (a) }}$}   
\vspace{5mm}

$^{{\rm (a)}}$ {\it UERJ, Universidade do Estado do Rio de Janeiro, \\
Rua S\~{a}o Francisco Xavier 524, 20550-013 Maracan\~{a}, Rio de Janeiro, Brazil.}

\vspace{3mm}

$^{{\rm (b)}}$ {\it Dipartimento di Fisica, Universit\'{a}
di Genova, \\ 
Via Dodecaneso 33 - I-16146 Genova, Italy - and INFN, Sezione di Genova.}

\vspace{3mm}

$^{{\rm (c)}}$ {\it Centro Brasileiro de Pesquisas F\'\i sicas (CBPF), 
\\Rua Dr. Xavier Sigaud 150 - 22290-180 - Rio de Janeiro - RJ - Brazil.}

\vspace{3mm} 

$^{{\rm (d)}}$ {\it Dipartimento di Fisica, Universit\'{a}
di Roma \ ''Tor Vergata '', 
\\Via della Ricerca Scientifica 1, 00173 Roma, Italy.}

\end{center}

\vspace{1cm}

It is known that supersymmetry affects in a deep way the
ultraviolet behaviour of quantum field theories leading, in the case of massless 
four-dimensional $N=2$ and $N=4$ super Yang-Mills (SYM), to a set of
remarkable nonrenormalization theorems for the corresponding $\beta _g$%
-functions [1]. 

In the case of $N=2,$ $\beta _g$ receives only one-loop contributions, while
in $N=4$ it vanishes to all orders of perturbation theory. The proof of these 
theorems is reviewed within the algebraic BRST approach [2,3]. 

The quantum properties of $N=2$ and $N=4$ SYM can be derived by making use 
of the twisting procedure, which allows to replace the spinor
indices of supersymmetry $(\alpha ,\dot{\alpha})$ with Lorentz vector
indices. The physical content of the theory is left
unchanged, for the twist is a linear change of variables, and the twisted
version is perturbatively indistinguishable from the original one. 

The proofs of the nonrenormalization theorems rely on the key
observation that the action of both $N=2$ and $N=4$ SYM is obtained from the
local gauge invariant polynomial ${\rm Tr}\phi ^2,$ $\phi $ being the scalar
field of the vector $N=2$ multiplet. This polynomial enjoys the important
property of having vanishing anomalous dimension, from which the 
nonrenormalization theorems for $N=2,4$ follow [2].

The global symmetry group of $N=2$
in four dimensional flat euclidean space-time is $SU(2)_L\times
SU(2)_R\times SU(2)_I\times U(1)_{{\cal R}}$, where $SU(2)_L\times SU(2)_R$
is the rotation group and $SU(2)_I$ and $U(1)_{{\cal R}}$ are the symmetry
groups corresponding to the internal $SU(2)$-transformations and to
the ${\cal R}$-symmetry. The twisting procedure consists of replacing the
rotation group by $SU(2)_L^{\prime }\times SU(2)_R$, where $SU(2)_L^{\prime }$ 
is the diagonal sum of $SU(2)_L$ and $SU(2)_I$, allowing to identify the
internal indices with the spinor indices. The fields of the $N=2$ vector
multiplet in the WZ gauge are given by $(A_\mu ,\psi _\alpha ^i,\overline{%
\psi }_{\dot{\alpha}}^i,\phi ,\overline{\phi })$, where $\psi _\alpha ^i,%
\overline{\psi }_{\dot{\alpha}}^i$ are Weyl spinors with $i=1,2$ being the
internal index of the fundamental representation of $SU(2)_I,$ and $\phi ,%
\overline{\phi }$ are complex scalars; all fields belonging to the adjoint
representation of the gauge group. Under the twisted group, these fields
decompose as [2] $ A_\mu \rightarrow A_\mu , \,\,\,\,\,\,
(\phi, \overline{\phi}) \rightarrow (\phi, \overline{\phi}) , \,\,\,\,\,\,
\psi^{i}_{\alpha}  \rightarrow   (\eta,\;\chi_{\mu \nu}) , \,\,\,\,\,\,
\overline{\psi}^{i}_{\dot{\alpha}} \rightarrow \psi_{\mu} $.

The fields $(\psi _\mu ,\chi _{\mu \nu },\eta )$ anticommute due to their
spinor nature, and $\chi _{\mu \nu }$ is a self-dual tensor field. The
action of $N=2$ SYM in terms of the twisted variables is [2] 
\begin{eqnarray}
&&S^{N=2} = \frac 1{g^2}{\rm Tr}\int d^4x\left( \frac 12F_{\mu \nu
}^{+}F^{+\mu \nu }+\frac 12\overline{\phi }\left\{ \psi ^\mu ,\psi _\mu
\right\} -\chi ^{\mu \nu }(D_\mu \psi _\nu -D_\nu \psi _\mu )^{+}\right.  
\nonumber \\
&&\left. \hspace{2mm}+\eta D_\mu \psi ^\mu -\frac 12\overline{\phi }D_\mu D^\mu \phi
-\frac 12\phi \left\{ \chi ^{\mu \nu },\chi _{\mu \nu }\right\} -\frac
18\left[ \phi ,\eta \right] \eta -\frac 1{32}\left[ \phi ,\overline{\phi }%
\right] \left[ \phi ,\overline{\phi }\right] \right) ,  \nonumber
\end{eqnarray}
where $g$ is the coupling constant and $F_{\mu \nu }^{+}=F_{\mu \nu }+\frac 12\epsilon _{\mu \nu \rho \sigma}F^{\rho \sigma }$.

Concerning the supersymmetry generators $(\delta _i^\alpha ,\overline{\delta 
}_{\dot{\alpha}}^i)$ of the $N=2$ superalgebra, the
twisting procedure gives rise to the twisted generators: a scalar $%
\delta $, a vector $\delta _\mu $ and a self-dual tensor $\delta _{\mu \nu }$%
, which leave the action invariant. Only the generators $\delta $ and 
$\delta _\mu $ are relevant for the nonrenormalization theorems [2].

The theory is quantized by collecting all the generators $%
(s,\delta ,\delta _\mu )$ into an extended operator $Q \equiv s+\omega \delta +\varepsilon ^\mu \delta _\mu$, $s$ being the BRST operator of the gauge transformations and $\omega $ and $\varepsilon ^\mu $ are global ghosts. 

>From the Batalin-Vilkovisky procedure, the complete gauge-fixed
action is $\Sigma =S^{N=2}+S_{{\rm gf}}+S_{{\rm ext}} $,  
where $S_{{\rm gf}}$ is the gauge-fixing term in the Landau gauge and $S_{%
{\rm ext}}$ contains the coupling of the non-linear transformations $Q\Phi _i
$ to antifields $\Phi _i^{*}$ [2]. 
The complete action $\Sigma $ satisfies the Slavnov-Taylor identity 
\begin{eqnarray}
{\cal S}(\Sigma )\equiv {\rm Tr}\int d^4x\left( \frac{\delta \Sigma }{\delta \Phi
_i^{*}}\frac{\delta \Sigma }{\delta \Phi _i}+b\frac{\delta \Sigma }{\delta 
\bar{c}}+\omega \varepsilon ^\mu \partial _\mu \bar{c}\frac{\delta \Sigma }{%
\delta b}\right) = \omega \varepsilon ^\mu \Delta _\mu ^{{\rm cl}}, \nonumber
\end{eqnarray}
where $ \bar c$ and $b$ are respectively the antighost and the Lagrange multiplier, and $\Delta _\mu ^{{\rm cl}}$ is an integrated local polynomial which is linear in the 
quantum fields and hence not affected by the quantum corrections [2].

>From the Slavnov-Taylor identity it follows that the linearized operator $%
{\cal S}_\Sigma $ 
\begin{eqnarray}
{\cal S}_\Sigma ={\rm Tr}\int d^4x\left( \frac{\delta \Sigma }{\delta \Phi
_i^{*}}\frac \delta {\delta \Phi _i}+\frac{\delta \Sigma }{\delta \Phi _i}%
\frac \delta {\delta \Phi _i^{*}}+b\frac \delta {\delta \bar{c}}+\omega
\varepsilon ^\mu \partial _\mu \bar{c}\frac \delta {\delta b}\right) 
\nonumber
\end{eqnarray}
is nilpotent modulo a total space-time derivative, namely  
${\cal S}_\Sigma {\cal S}_\Sigma =\omega \varepsilon ^\mu \partial _\mu $. 

Introducing the operator ${\cal W}_\mu $
\begin{eqnarray}
{\cal W}_\mu \equiv \frac 1\omega \left[ \frac \partial {\partial \varepsilon ^\mu
},{\cal S}_\Sigma \right], \;\;\;\;\;\;\ 
\left\{ {\cal W}_\mu ,{\cal S}_\Sigma \right\} =\partial _\mu
,\;\;\;\;\;\;\left\{ {\cal W}_\mu ,{\cal W}_\nu \right\} =0,  \nonumber
\end{eqnarray}
it follows  that the complete action $\Sigma $ is obtained by repeated applications of $%
{\cal W}_\mu $ on ${\rm Tr}\phi ^2$, according to the formula [2]
\begin{eqnarray}
\frac{\partial \Sigma }{\partial g}=\frac 1{3g^3}\epsilon ^{\mu \nu \rho
\sigma }{\cal W}_\mu {\cal W}_\nu {\cal W}_\rho {\cal W}_\sigma \int d^4x%
\;{\rm Tr} \phi ^2 \;\;{\rm +}\;\;{\cal S}_\Sigma \Xi ^{-1}{\rm ,}
\nonumber
\end{eqnarray}
for some irrelevant trivial cocycle ${\cal S}_\Sigma \Xi ^{-1}$. This
equation is of fundamental importance for the nonrenormalization theorems. It states that 
the bulk of the theory is the
composite operator ${\rm Tr}\phi ^2$, which  contains all the
information on the ultraviolet behaviour. This is a consequence of the fact that the operator ${\rm Tr}\phi ^2$ obeys 
a Callan-Symanzik equation with vanishing anomalous
dimension [2].
It follows thus that the $\beta _g-$function obeys the 
differential equation [2] 
\begin{eqnarray}
 g\frac{\partial \beta _g}{\partial g}- 3\beta _g=0\; \;\;\;\;\; \Rightarrow  \;\;\;\;\;
\beta _g={k\,}g^3,\;\;\;\;({k\; \rm constant)\;.} \nonumber 
\end{eqnarray}
This equation expresses the celebrated nonrenormalization theorem of $N=2$
SYM, stating that, to all orders of perturbation theory, the $\beta _g-$%
function has only one-loop contributions.

The above procedure has been successfully extended to the presence of matter, with the hypermultiplets belonging to a generic representation  of the gauge group. Also in that case, a complete twisted formulation of the theory, namely action, fields and symmetries, has been given. 

A direct one-loop computation of $k$ yields [4] 
\begin{eqnarray}
k=-\frac{1}{8\pi ^2}\left( C_1-H C_2\right) \;, \nonumber
\end{eqnarray}
where $C_1$ and $C_2$ are respectively the Casimir invariants of the
representations of gauge and matter $N=2$ multiplets, and $H$ is the number
of matter multiplets.  The $N=4$ is recovered as a particular case of $N=2$, with matter in the adjoint representation of the gauge group and $H=1$. It remains true that the fully quantized 
twisted action of $N=4$ is related to the invariant polynomial $\rm{Tr }\phi^2$ [3]. This implies that the beta function of $N=4$ can be at most of one-loop order. Therefore, from $H=1$ and $C_1=C_2$, it vanishes to all orders of perturbation theory.

\vspace{4mm}

{\bf BIBLIOGRAPHY.}

[1] P. West, "Introduction to Supersymmetry and Supergravity", Singapore, {\it Singapore: World 
Scientific} (1990).

[2] A. Blasi, V.E.R. Lemes, N. Maggiore, S.P.Sorella, A. Tanzini, O.S. Ventura and  
L.C.Q. Vilar, JHEP 0005 (2000) 419 (hep-th/0004048).

[3] V.E.R. Lemes, M.S. Sarandy, S.P. Sorella, A. Tanzini and O.S. Ventura, (hep-th/0011001).

[4] S. Weinberg, "The Quantum Theory of Fields, Vol.III Supersymmetry", Cambridge Univ. Press, 2000.

\enddocument